# Supersaturation model for InN plasma-assisted MOCVD

Z. Ahmad, M. Vernon, G.B. Cross, D. Deocampo, and A. Kozhanov

Department of Physics and Astronomy and Center for Nano-Optics, Georgia State University, Atlanta, GA 30303

We developed a thermodynamic supersaturation model for plasma-assisted metalorganic chemical vapor deposition of InN. The model is based on the chemical combination of indium with plasma-generated atomic nitrogen ions. Indium supersaturation was analyzed for InN films grown by PA-MOCVD with varying input flow of indium precursor. Raman spectroscopy, X-ray diffraction, and atomic force microscopy provided feedback on structural properties and surface morphology of grown films. Growth parameter variation effect on In supersaturation was analyzed. InN films grown at varying growth parameters resulting in the same In supersaturation value exhibit similar structural properties and surface morphology.

The III-nitrides (AlN, GaN, InN) are prominent direct bandgap semiconductor materials well known for their optoelectronic and power electronics applications[1,2,3]. InN has a bandgap energy of 0.7eV[4], which makes it a promising material for applications in infrared optoelectronic devices, including emitters, sensors, and tandem solar cells[5]. InN based nanostructures[6] can also be used in devices operating in the THz spectral range.[7]

InN growth has been the focus of many studies in the past several decades. InN growth is typically limited to lower growth temperatures due to the low InN dissociation temperature[42] and higher nitrogen equilibrium pressure over the grown InN film surface[43]. The growth of InN at low temperatures (~500 °C), further restricts the metalorganic vapor phase epitaxy (MOVPE) and high pressure chemical vapor deposition (HPCVD)[1] growth of InN due to the low decomposition rate of ammonia. The problem of the low number density of active nitrogen species can be addressed by using nitrogen plasma as a nitrogen precursor. Plasma-assisted MBE (PA-MBE)[12,13,14,15] and plasma-enhanced atomic layer deposition (ALD)[16] have demonstrated epitaxial InN growth using nitrogen plasma. Recent studies indicate that plasma-assisted MOCVD growth of InN allows growing InN at temperatures above its decomposition temperature[44]. This opened possibilities of InN growth within GaN and AlN growth temperature window, thus allowing for InN-GaN-AlN based heterostructures InN nano-pillar and wire growth have been done by PA-MOCVD at low temperatures[17, 18]. At the same time, InN grown by PA-MOCVD is still not well studied, and epitaxial atomically-flat film growth required for heterostructures is not achieved in this system yet.

PA-MOCVD growth of InN is controlled by several process parameters such as the substrate temperature, reactor pressure, V/III ratio, diluent gas, plasma power, and input partial pressures of indium-metal and nitrogen species[30]. With that many parameters, it is a known challenge to find a suitable growth parameter window and distinguish the influence of each parameter on the growth and properties of resulting thin films.

A comprehensive model of the PA-MOCVD growth process of InN that incorporates all process parameters would be helpful in the prediction of optimal growth conditions. The thermodynamics of the growth process provides an understanding of the forces which drive crystal growth. A model based on thermodynamic supersaturation and chemical potential could provide a simplified method of InN growth optimization. Conventionally, the MOCVD growth of III-N is carried out under III-species mass-transport limitation[45,46,47]. As a result, the supersaturation for InN growth at given conditions can be obtained by normalizing the difference in the input partial pressure of indium and the equilibrium vapor pressure of In above the InN growth surface. According to the Burton–Cabrera–Frank (BCF) theory[19], the behavior of the growth surface is strongly dependent on the supersaturation, that is when the crystal is growing under a supersaturated environment.

Previous studies indicate that the incorporation of the III-metal vacancy is strongly sensitive to the growth stoichiometry during the III-N growth.[20] For a conventional MOCVD process at a constant substrate temperature, the growth stoichiometry seems to be easily controlled by varying the V/III ratio. Lower V/III ratio results in more III-rich conditions. However, this idea gives conflicting outcomes. For example, in work by Saarinen et al.[21], the Ga vacancies measured by positron annihilation spectroscopy significantly increased from $10^{16}$ to $10^{19}$ cm$^{-3}$ with the increase of V/III ratio from 1000 to 10,000. However, Xin et al[22] report that higher V/III ratio growth conditions did not exhibit a similar increase in Ga vacancy concentration.

For the PA-MOCVD growth, the use of thermodynamic supersaturation (and hence chemical potential) could prove to be more convenient in evaluating the consequence of III-metal rich or N-rich growth conditions, rather than typically used V/III ratio.

In this letter, the role of the supersaturation in the PA-MOCVD InN growth is studied. A method for determining supersaturation as a function of growth parameters such as input flows and RF power is presented. The surface morphology and the structural properties of grown InN films were found to be strongly correlated with In supersaturation.

Two InN film sample sets (labeled as A and B) were grown using PA-MOCVD on c-plane Al$_2$O$_3$ wafers offcut at 0.2° towards m-plane. Trimethylindium (TMI) and nitrogen plasma were used as group-III and group-V precursors, respectively. The nitrogen plasma was produced using an RF hollow cathode plasma source. Nitrogen gas was used as the TMI carrier gas. The sapphire wafers were cleaned in hydrogen gas at 250°C in the growth chamber. An InN buffer layer was deposited at 550°C. The sample was then heated to the growth temperature of 775°C. For the sample set A, four samples were grown with TMI input flows varying from 8.2 µmol/min to 12.1 µmol/min, for which the calculated In supersaturation ranges from 4.21 to 8.26. The input N$_2$ flow through the plasma source was fixed at 336 µmol/min, the RF power was fixed at 150 W, and the reactor pressure at 1.65 Torr. For sample set B, three samples were grown at varying TMI input flows, reactor pressure, and RF power calculated to have the same In supersaturation of 2 and hence the same chemical potential. The process parameters for samples in the set B were: 2.4 Torr, 400 W, V/III=194 (sample 1), 3.2 Torr, 530 W, V/III=217 (sample 2), and 3.8 Torr, 550 W, V/III=258 (sample 3). Room temperature Raman spectroscopy in backscattering geometry was used to investigate the structural composition of grown films. Phonon mode positions and FWHM were extracted using a multi-peak fit of the experimentally acquired Raman spectra for the samples in this study. The growth rate was determined by fitting interference fringes measured by Fourier-transform infrared (FTIR) spectroscopy.

The chemical reaction forming solid-phase InN is driven by the supersaturation of In species in the gas phase. Thin-film growth processes, including PA-MOCVD, are non-equilibrium processes. The driving force for thin film formation is the deviation from the thermodynamic equilibrium and can be quantitatively described by the change in free energy of the process $\Delta G = -RT \ln(1 + \sigma)$[23], where $\sigma$ is the supersaturation of indium defined by Eq. (1).

$$\sigma = \frac{P_{In}^{input} - P_{In}^{eq}}{P_{In}^{eq}} \quad (1)$$

Here $P_{In}^{input}$ is the input partial pressure of indium and $P_{In}^{eq}$ is the equilibrium partial pressure of indium at given reaction conditions. The supersaturation of In in the present work was calculated following the procedure described by Koukitu et al. for GaN[24]. The chemical reaction in the reactor can be described by $In + \frac{1}{2}N_2 \overset{RF}{\Leftrightarrow} InN$. The equilibrium constant corresponding to this reaction can be written as $K = a_{InN} / P_{In} P_{N_2}^{1/2}$, where $a_{InN}$ is the activity of InN, $P_{In}$ is the equilibrium vapor pressure of In, $P_{N_2}$ is the partial pressure of $N_2$, and, $\log_{10} K(T) = 15.7 + 3.82 \times \frac{10^4}{T} + 0.289 \log_{10}(T)$ where T is the temperature[25,26,27,28,29]. The total pressure of the system assumed constant during the process can be written as

$$P_T = P_{In}^{eq} + P_{plasma\ N_2} + P_{IG} \quad (2)$$

Where IG stands for "inert" gases that are not contributing to the final product. The reduced total pressure $P_r$ involves only the partial pressures of species included in the chemical reaction and hence is written in as

$$P_r = P_T - P_{IG} = P_{In}^{input} + P_{plasma\ N_2}^{input} \quad (3)$$

As described in our previous work[30], the chemical reaction occurring at the growth surface resulting in solid InN formation is defined by the chemical combination of atomic nitrogen ions $N^*$ with indium. We introduce the parameter $\gamma$ is the ratio of number atomic nitrogen ions to the total number of in-flowing nitrogen molecules to account for a fraction of atomic nitrogen ions $N^*$ generated by the plasma source from neutral molecular nitrogen. $\gamma$ is a function of the reactor pressure and RF power for given nitrogen flow through the plasma source. The plasma dissociation of $N_2$ into ionized atomic nitrogen and other neutral and ionized molecular nitrogen species can be written as:

$$N_2 \xrightarrow{R.F} 2\gamma N^* + (1-\gamma) N_2 \quad (4)$$

For any given PA-MOCVD growth process, $\gamma$ can be determined by analyzing the in-situ optical emission spectra of nitrogen, using atomic emission spectroscopy[31,32,33,34,35]. The Saha–Boltzmann equation was utilized to analyze the nitrogen plasma composition[36]. The detailed method of the plasma analysis used for PA-MOCVD of InN is discussed in our previous work[30]. Calculated dependence of $\gamma$ on reactor pressure and plasma RF power is shown in Figure 1.

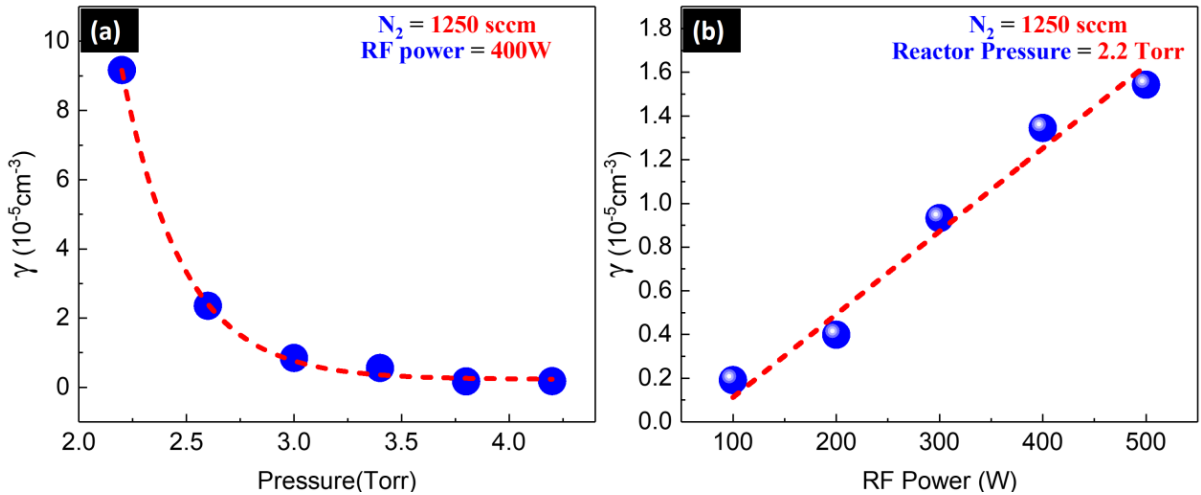

**Figure 1:** (Color online) $\gamma$ the ratio of the concentration of atomic nitrogen ions to the input nitrogen, as a function of reactor pressure **(a)** and RF power **(b).**

Following Eq.(4), the reduced pressure in Eq.(3) can be rewritten as

$$P_r = P_{In}^{input}\left\{1 + \frac{V}{III}(1+\gamma)\right\} \quad (5)$$

where $\frac{V}{III}$ is the ratio of the input flow of nitrogen to the input flow of indium. The molar conservation considerations give the following relation between partial pressures can be expressed in Eq. (6).

$$P_{In}^{input} - P_{In}^{eq} = P_{plasma\,N_2}^{input} - P_{plasma\,N_2}^{eq} \quad (6)$$

We solve equations 3, 5 and 6 along with $K = \frac{a_{InN}}{P_{In}P_{N_2}^{1/2}}$, to find a polynomial equation (Eq. (7)) for equilibrium partial pressure of indium $P_{In}^{eq}$.

$$(P_{In}^{eq})^3 - \left\{3P_{In}^{input} + (1+\gamma)P_{plasma\,N_2}^{input}\right\}(P_{In}^{eq})^2 + \left(\frac{a_{InN}}{K}\right)^2 = 0 \quad (7)$$

Eq.(7) can be solved for partial equilibrium pressure of Indium for any growth conditions used for PA-MOCVD of InN. Furthermore, substituting the solution of the Eq. (7) into Eq. (1) will give the indium supersaturation value for a given growth process parameter set.

The role of the other nitrogen plasma radicals such as atomic neutrals and molecular ions is not considered due to evidence that their influence on InN growth is insignificant[30]. Consequently, the reduced total pressure (Eq.5), which remains constant throughout the growth process, is defined in terms of the input partial pressure of group III species, input V/III ratio, and the parameter $\gamma$ that can be determined from in-situ optical emission spectra[30]. The change in Gibbs free energy of the III-metal can be found as $\Delta G = -k_B T \ln(1+\sigma)$ which is equal to the chemical potential per mole of III-metal[23].

Various growth parameter combinations can lead to similar supersaturation and chemical potential values. This should result in the growth of InN films exhibiting similar structural properties, which might not be evident from standing alone growth parameter combinations. Effect of the reactor pressure, substrate temperature, and TMI flow during growth on indium supersaturation $\sigma$ calculated following the thermodynamic model described above is shown in Figure 2, with the molar flow of indium reasonably having the most significant effect. The results of these calculations were used to choose the growth parameter set values. TMI flow variation was used to control the In supersaturation in the sample set A. Combined variations of the TMI flow, reactor pressure, and plasma power were used to set the In supersaturation values equal 2 in all set B samples.

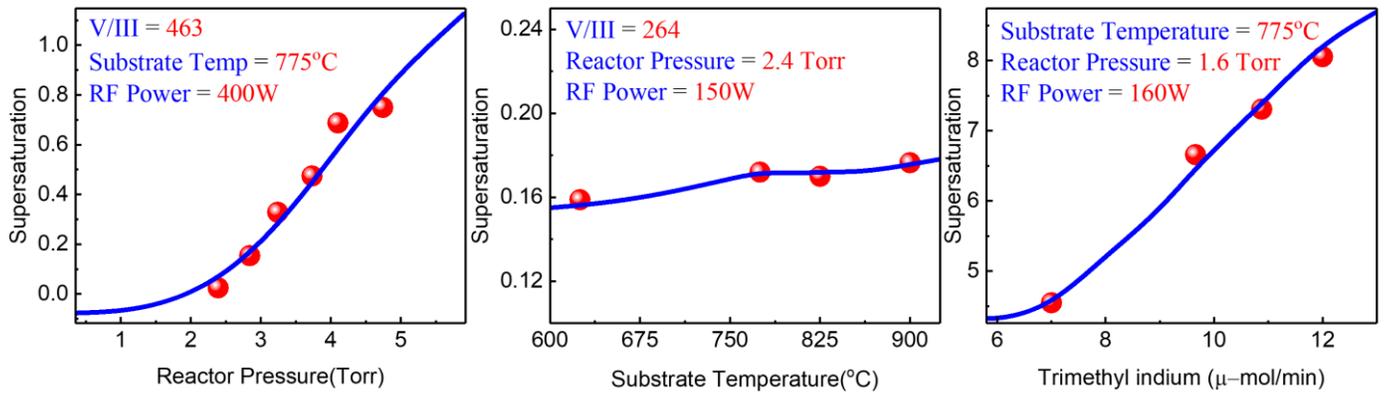

**Figure 2** (Color online)**:** Dependence of In supersaturation on the reactor pressure (left), substrate temperature (center), and TMI flow (right) calculated for PA-MOCVD. Model calculation results are shown as solid lines. Actual growth parameter values used in the growth and corresponding supersaturation values are shown as dots.

The influence of TMI flow controlled In supersaturation on the structural properties of the grown films has been studied via Raman spectroscopy. Raman spectra for the two sets of InN samples grown at various supersaturation values are shown in Fig.3.

The InN $E_2$-high ( 489 cm$^{-1}$ ) and $A_1$-LO ( 590 cm$^{-1}$ ) phonon modes were observed in each of the sample sets. The phonon relaxation time was estimated using the FWHM and peak position of the $E_2$-high and $A_1$-LO phonon modes extracted from a multi-peak fit of the Raman spectra[37]. Fig.4 shows the $E_2$-high phonon mode relaxation time as a function of supersaturation for both sample sets A and B. The $E_2$-high phonon relaxation time decreases with the increase of indium supersaturation in sample set A. It is likely caused by the increase in nitrogen vacancies in the grown InN, which are primary contributors to the $E_2$-high phonon mode[37]. No significant correlation has been observed between the $A_1$-LO phonon mode relaxation time and the indium supersaturation. For the sample set B ($\sigma = 2$), the $E_2$-high phonon relaxation time remains the same for the whole series. The indium supersaturation is kept almost constant for all samples in the set B. This results in the same chemical potential and, hence, dictates the similar reaction kinetics. As a result, the grown films exhibit similar structural properties.

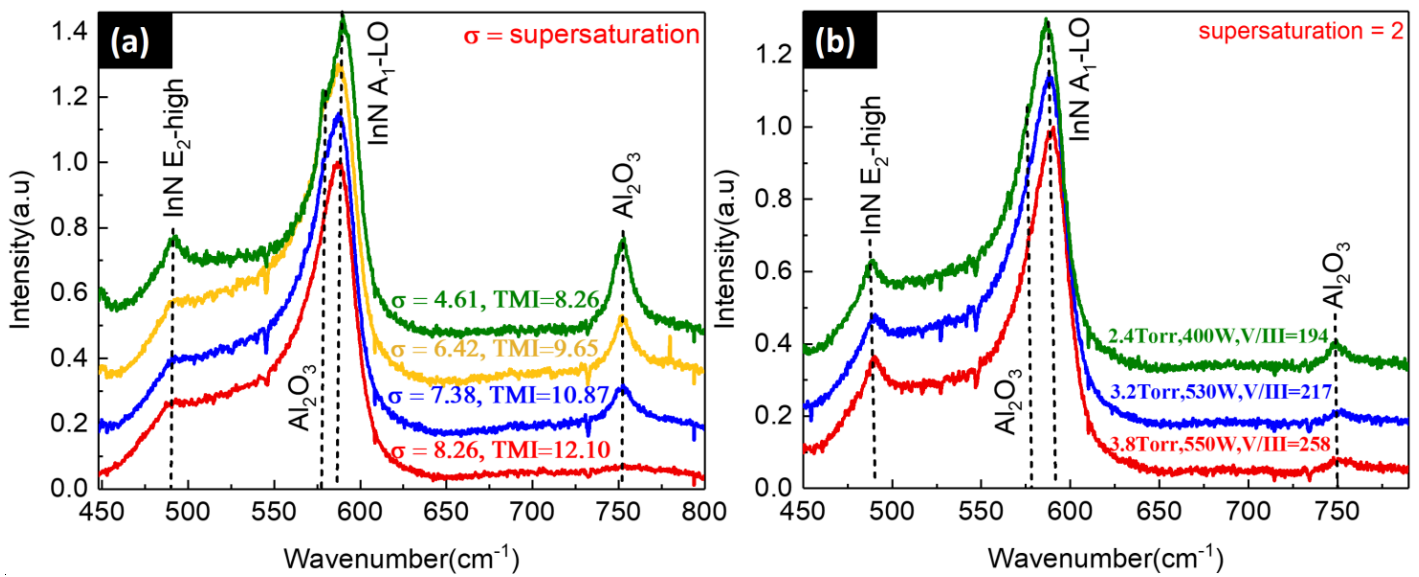

**Figure 3:** (Color online) **(a)** Raman spectra measured on Set A InN films with varying TMI flow. Numbers indicate the TMI flow and indium supersaturation levels. **(b)** Raman spectra measured on Set B InN films. Numbers indicate the growth conditions resulting in $\sigma = 2$.

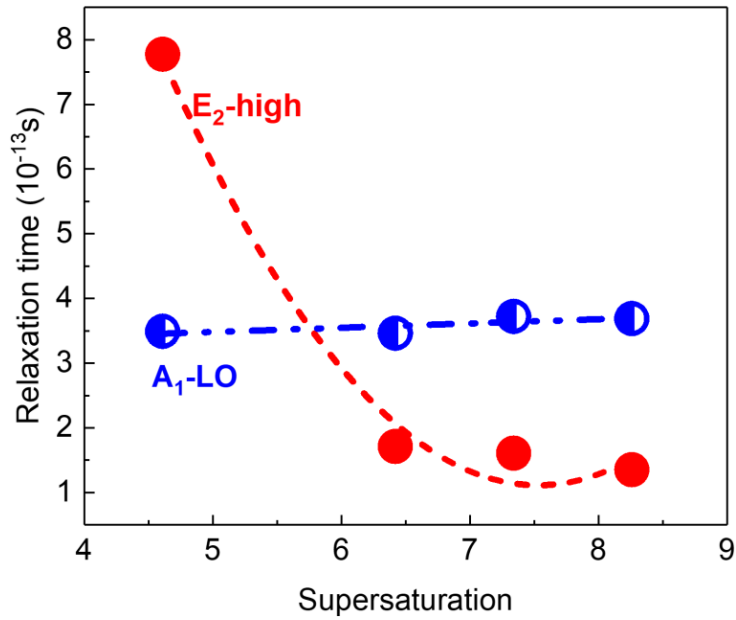

**Figure 4:** (Color online) The $E_2$-high and $A_1$-LO phonon mode relaxation time dependence on indium supersaturation level measured for set A InN films.

XRD spectra were recorded on several samples from Set B. Figure 5 shows $\theta$-$2\theta$ scans for these samples. For all analyzed samples, peaks located at 31.38º, 64.90º, and 41.85º were identified as (0002) InN, (0004) InN, and (0006) $Al_2O_3$ substrate respectively. Lower intensity peaks of $(10\bar{1}1)$ InN and $(20\bar{2}2)$ InN located at 33.0º and 69.22º, respectively, indicate the polycrystalline structure of the grown film. To achieve monocrystalline film, most likely, the InN growth initiation on the sapphire surface should be optimized. The obtained XRD results are consistent with results previously reported by Gao et al.[38] for InN grown at 550 ˚C by MBE. Peaks labeled as "*" are likely artifacts of the 2º-offcut c-plane sapphire substrate, as described by Mukherjee et al[39]. The crystallite size in each of the samples in the set B was determined using Scherrer equation[40,41] given as

$\tau = k\lambda/\beta sin\theta_{hkl}$, where $\tau$ is the crystalline domain size, $\lambda$ is the X-ray wavelength, $\beta$ is FWHM of the corresponding XRD peak and $\theta_{hkl}$ is the diffraction angle. The crystallite size equal to 73±3 nm was calculated for all analyzed set B samples.

The similarity of structural properties of samples grown at the same In supersaturation level seen in the Raman spectra and XRD is complemented by the similar behavior of grain sizes determined by AFM. Figure 6 shows 1x1µm² AFM micrographs of InN films grown at widely different growth conditions resulting in σ=2. The surfaces of these films exhibited similar surface morphologies with RMS roughness of (4.34±0.07)nm and average grain size of (42.5±2.5)nm.

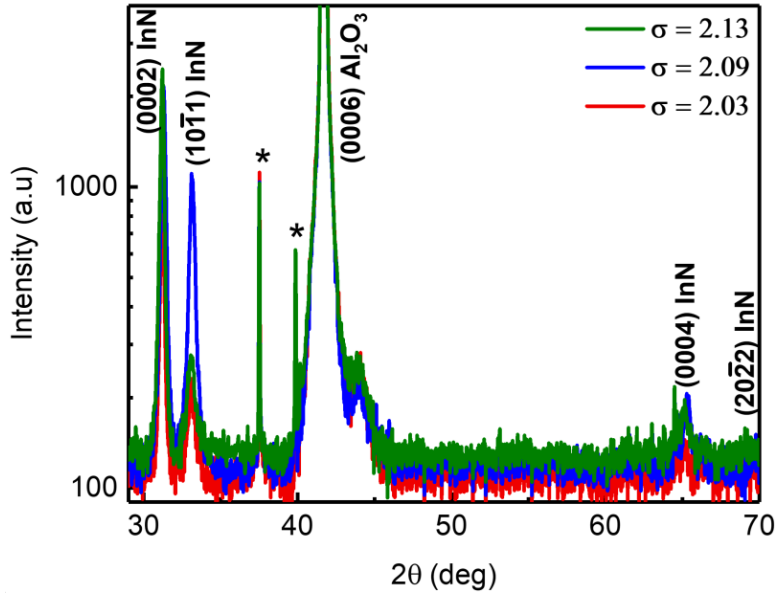

**Figure 5:** (Color online) XRD spectra measured on InN films grown at widely different conditions, all of which result in an indium supersaturation value of approximately 2.0.

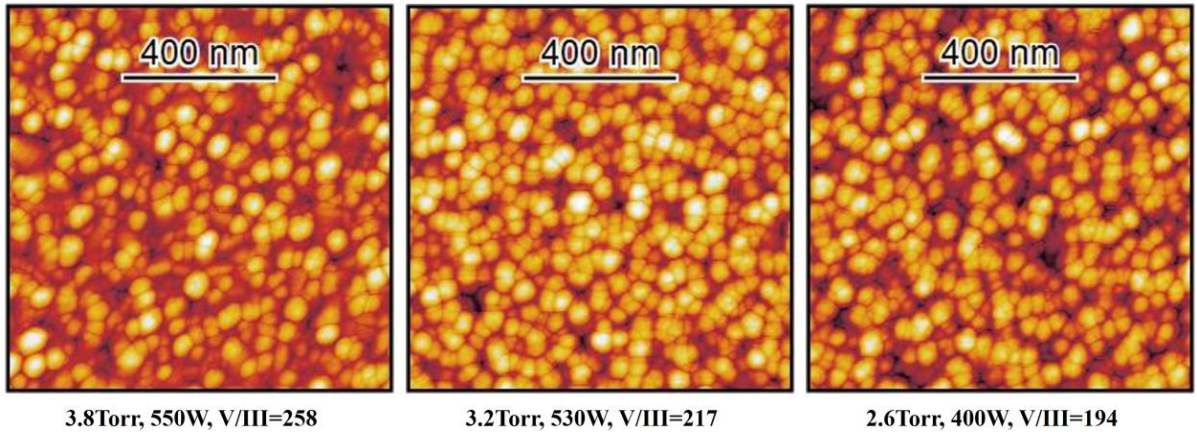

**Figure 6:** (Color online) Surface morphology measured on InN films plotted for set B via Atomic Force Microscopy. Numbers indicate the growth conditions resulting in indium supersaturation level of 2.

In summary, we applied the indium supersaturation model to the PA-MOCVD process, in which ionized atomic nitrogen ions are used as group V precursor. The developed model allowed calculation of the supersaturation as a function of growth parameters such as reactor pressure, plasma power, indium precursor flow, and substrate temperature. The effect of the input TMI was observed to be most significant. The model also allowed finding multiple growth parameter combinations that result in the same value of supersaturation. InN grown at the same In supersaturation level demonstrated similar structural properties studied via Raman spectroscopy and XRD, and similar surface morphology studied via AFM. Supersaturation or chemical potential parameters can be used as predictive feedback for the InN and other materials growth in plasma-assisted MOCVD and should aid in achieving desired film qualities.

This work was supported by the DOE Grant No. NA-22- WMS-#66204 via PNNL subcontract and NSF Grant No. EAR1029020. The authors are thankful to Ahmal J. Zafar for helpful discussions and to Pete Walker for help with the growth facility maintenance.